\documentclass[preprint,prd,amsmath,amssymb,amsthm,nofootinbib]{revtex4}
\usepackage[mathscr]{euscript}
\usepackage{bm}
\usepackage{graphicx}
\usepackage{subfigure}
\usepackage{overpic}
\usepackage[colorlinks=true,linkcolor=red]{hyperref}
\newcommand{\bea}{\begin{eqnarray}}
\newcommand{\eea}{\end{eqnarray}}
\newcommand{\beq}{\begin{equation}}
\newcommand{\eeq}{\end{equation}}

\def\/{\over}

\begin{document}
\title{Production of gravitational waves during preheating with nonminimal coupling}

\author{ Chengjie Fu$^{1}$,  Puxun Wu$^{1,2}$\footnote{Corresponding author: pxwu@hunnu.edu.cn} and Hongwei Yu$^{1}$\footnote{Corresponding author: hwyu@hunnu.edu.cn}  }
\affiliation{$^1$Department of Physics and Synergetic Innovation Center for Quantum Effects and Applications, Hunan Normal University, Changsha, Hunan 410081, China \\
$^2$Center for High Energy Physics, Peking University, Beijing 100080, China }

\begin{abstract}
We study the preheating and the in-process
production of gravitational waves (GWs) after inflation  in which  the inflaton is nonminimally coupled to the curvature in a self-interacting quartic potential  with the method of lattice simulation. We find that the  nonminimal coupling enhances the  amplitude of the density spectrum of inflaton quanta, and as a result,  the peak value of the GW spectrum generated during preheating  is  enhanced as well  and
 might reach the limit of detection in  future GW experiments.    The peaks of the GW spectrum not only exhibit    distinctive characteristics as compared to those of minimally coupled inflaton potentials but also imprint information on the nonminimal coupling and the parametric resonance, and thus the detection of these peaks in the future will provide us a new avenue to reveal the physics of  the early universe.

\end{abstract}


\maketitle
\section{Introduction}
 Inflation, an accelerated expansion in the early universe, is an elegant idea proposed to resolve  the horizon, flatness, and monopole problems in the big bang standard cosmology~\cite{Inflation}. At the same time, the quantum fluctuations  of the inflaton field  also provide the seed for the formation of cosmic structures~\cite{Fluc}.  The slow-roll single-field inflationary models predict that the fluctuant  spectrum of  curvature  perturbations is  nearly scale-invariant. This prediction is consistent with the  observations of  the cosmic microwave background (CMB)~\cite{COBE, WMAP}, which limit  the spectral index to be $n_s=0.968\pm 0.006$ at $68\%$ confidence level (CL)~\cite{Planck}.

During inflation, there are also  tensor perturbations of the spacetime metric, which lead to the production of a stochastic background of gravitational waves (GWs).  Since the amplitude of the power spectrum of GWs depends on the energy scale of inflation in the  slow-roll single-field inflationary models, the combination of the ratio of tensor to scalar fluctuations $r$, which is constrained to be $r<0.09$ by current CMB data~\cite{Planck}, and the spectral index $n_s$ is capable of discriminating  a host of inflationary models. For example,  the simple cubic and quartic potentials are nearly ruled out by the Planck 2015 data, and the simple quadratic potential is also disfavored~\cite{Planck}.

After inflation, the universe enters a reheating era in which   the potential energy of the inflaton is transferred  to a thermal bath of the matter species that are present in our Universe today. The first stage of reheating is preheating~\cite{Traschen, Bassett}, in which there exists an explosive particle production of the inflaton quanta or a scalar matter field coupled to the inflaton due to the parametric resonance. Since only a part of momenta  are in the resonance bands, the Fourier modes of the inflaton quanta or the scalar matter field with resonant  momenta  grow exponentially while all other modes do not. This  results in that  the matter distribution has large and time-dependent density inhomogeneities  in the position space, and thus possesses  substantial quadruple moments. Therefore, the parametric resonance of preheating can source a significant production of GWs~\cite{Kofman94, Khlebnikov}. Different from vacuum fluctuations of tensor perturbations during inflation, the amplitude of the GW spectrum generated during preheating is independent of the energy scale of inflation which only determines the present peak frequency~\cite{Easther, Easther2}. Recently, the production of GWs during preheating with special inflation potentials,  such as asymmetric potential around the minimum and cuspy potential, has  been investigated in~\cite{Antusch, Liu}. It was found that there is a pronounced peak in the GW spectrum  for the asymmetric potential,  and the pronounced peak becomes two in the case of the cuspy potential.

Although simple quadratic  and quartic potentials are  disfavored phenomenologically,  they  however agree with observations very well after  a simple extension which assumes  the existence of a nonminimal coupling between the  inflation field and the curvature scalar~\cite{Planck13, SJS}.  Moreover,  nonminimal couplings can be  generated naturally when quantum corrections are considered and are essential for the renormalizability of the scalar field theory in curved space~\cite{Callan}.  Therefore, it is of great interest to investigate the  production of GWs during preheating in the case of symmetric and simple power law inflationary potentials with nonminimal couplings as opposed to those potentials with rather peculiar shapes and see whether the nominimal coupling shows  as peculiar gravitational wave signatures.  To understand  the physics of the GW production in details, the process of preheating  needs to be investigated thoroughly.  However,  all current  studies (to the best of our knowledge) on the preheating after  inflation with nonminimal couplings consider only the linear perturbations of the scalar field and use the  Hartree approximation to  account for the backreaction of the amplified quantum fluctuations~\cite{Tsujikawa99, Tsujikawa00, Fu2017}, and this leaves unfortunately some other  interesting  physical  processes,  such as scattering among different modes and the evolution of the energy density spectra of inflaton quanta,  unclear. So, in this paper,  we first plan to fill this gap by a full investigation  of preheating after the nonminimally coupled scalar field inflation with a self-interacting potential by  performing a numerical lattice simulation. We find that the nonminimal coupling has an appreciable effect on the amplitude of the density spectrum of inflaton quanta,  since, as is revealed by our study, the amplitude grows with the increase of the coupling parameter.     Then we   study the production of GWs during preheating. The existence of the nonminimal coupling gives rise to two new source terms  in the GW equation, which become more and more important  as the coupling parameter grows. With the increase of the coupling parameter,  the amplitude of GW spectrum also grows constantly and  might
 reach the detection limit of future GW  experiments.

\section{Preheating}
In a flat Friedmann-Robertson-Walker (FRW) background, the scalar field $\phi$  nonminimally  coupled with  gravity in the form $\xi \phi^2 R$ satisfies the  equation
\begin{equation}\label{1}
\ddot\phi+3H\dot\phi-\frac{1}{a^2}\nabla^2\phi+\frac{dV(\phi)}{d\phi}+\xi R\phi=0\;,
\end{equation}
where $\xi$ is the coupling parameter,  $R$ is the curvature scalar,  a dot denotes the derivative with respect to the cosmic time $t$, $a$ is the cosmic scale factor,  $H=\frac{\dot{a}}{a}$ is the Hubble parameter, $V(\phi)=\frac{\lambda}{4} \phi^4$ is the self-interacting potential with $\lambda$ being a constant, and $\nabla^2$ is the  Laplacian operator. This inflationary model is consistent with observations since the negative nonminimal coupling yields a slight increase of $n_s$ as well as a significant decrease of $r$~\cite{Felice, SJS}. The joint of Planck and WMAP polarization data gives $\xi<-0.0019$ at  $95\%$ CL~\cite{Planck13, SJS}.  After inflation, the scalar field oscillates around $\phi=0$.

 We assume that the inflaton is weakly coupled to other fields and thus only the parametric resonance of inflaton quanta  is considered during preheating. For convenience, one can divide the scalar field into the homogeneous and fluctuant parts: $
\phi(t,\mathbf{x})=\phi_0(t)+\delta\phi(t,\mathbf{x})$. To the linear order, the perturbations in the momentum space obey the equation of motion
\begin{equation}\label{2}
\delta\ddot\phi_k+3H\delta\dot\phi_k+\biggl[\frac{k^2}{a^2}+3\lambda\phi_0^2+\xi  R\biggl]\delta\phi_k=0\;.
\end{equation}
By defining a conformal field $\varphi_0\equiv a\phi_0$ and its fluctuation $\delta\varphi_k\equiv a\delta\phi_k$ and introducing the conformal time $\eta\equiv\int a^{-1} dt$, Eq. (\ref{2}) can be rewritten as
\begin{equation}\label{3}
\frac{d^2}{d\eta^2}\delta\varphi_k+\omega_k^2\delta\varphi_k=0\;,
\end{equation}
 with
\begin{equation}\label{4}
\omega_k^2\equiv k^2+3\lambda\varphi^2_0+\left(\xi-\frac{1}{6}\right)R a^2\;.
\end{equation}
Since the background field  is  time dependent,  Eq.~(\ref{3}) describes an oscillator with a time varying frequency.  When $\xi=0$, due to that the evolution of the scale factor can be approximately expressed  as $a\sim \eta$ after using the time averaged relation $\frac{1}{2}\langle \dot{\phi}^2\rangle=2\langle V\rangle$, the time averaged value of $R$ vanishes as a results of $R=\frac{6}{a^3} a''$, where  a prime denotes the derivative with respect to the conformal time $\eta$.  Eq.~(\ref{3}) belongs to the class of Lam$\acute{e}$ equation, and thus $\delta \varphi_k$ has  exponentially growing modes for certain momenta $k$. But the resonant bands are in  narrow ranges since the amplitude of $\varphi_0$ is very small.
For the nonminimal coupling case,  the oscillations of the inflaton  and its perturbations differ as  compared to the minimal coupling case because of the last term in Eqs.~(\ref{1}) and (\ref{4}), so that  the resonant structure  is modified  and the  resonant bands become broad with the increase of $|\xi|$.

 To obtain a panoramic  view  of preheating,  the best way is to perform numerical lattice simulations. By modifying the publicly available C++ package LATTICEEASY \cite{GI},  we investigate the evolution of  the nonminimally coupled inflaton field   in the configuration space in an expanding universe. The results are simulated with $N^3=(128)^3$ points and the minimum momentum $k/(\sqrt{\lambda}\phi_i)\sim\mathcal{O}(0.2)$. Here,  $\phi_i$ is  the value of the scalar field at the beginning of   preheating, which is given by~\cite{Tsujikawa00, Fu2017}
\begin{equation}\label{6}
\phi_i=\left[\frac{\sqrt{(1-24\xi)(1-8\xi)}-1}{16\pi(1-6\xi)|\xi|}\right]^{1/2}m_{pl}\;,
\end{equation}
where $m_{pl}$ is the Planck mass. We  use the conformal vacuum as the initial state with the initial conditions of the fluctuations being $\delta \varphi_k(0)=1/\sqrt{2\omega_k(0)}$ and $\delta{\varphi}'_k(0)=-i \omega_k(0) \delta\varphi_k(0)$, and stop the simulation when the density spectrum of the inflaton quanta does not change appreciably.

\begin{figure}
\centering
 \includegraphics[width=0.49\textwidth ]{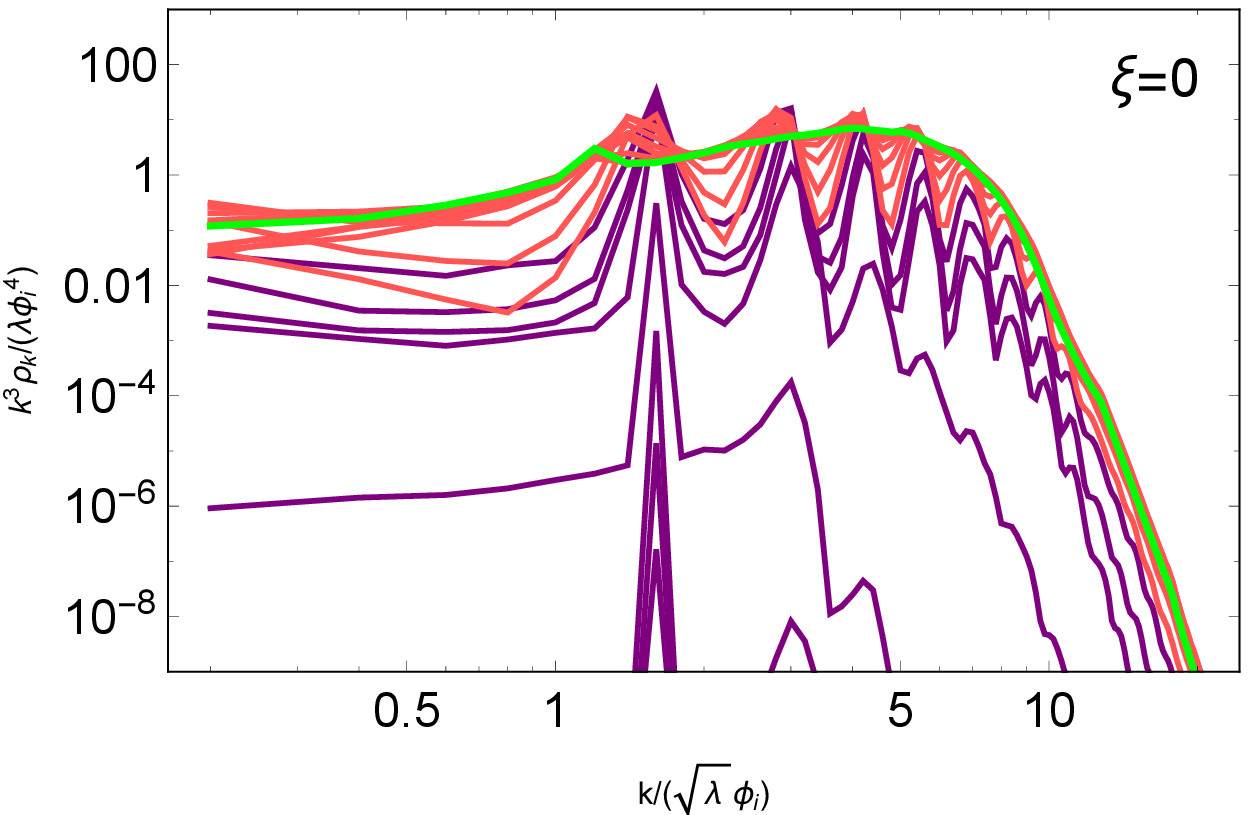}
 \includegraphics[width=0.49\textwidth ]{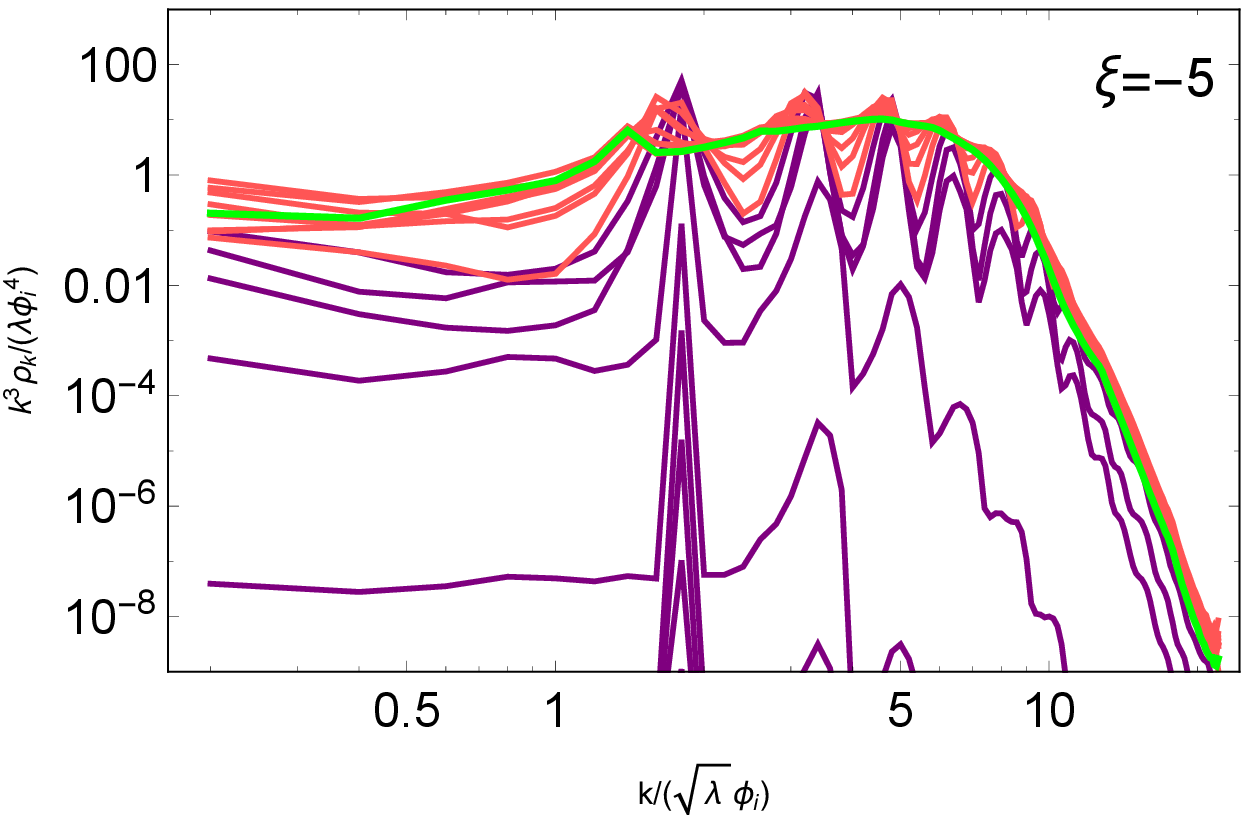}
 \includegraphics[width=0.49\textwidth ]{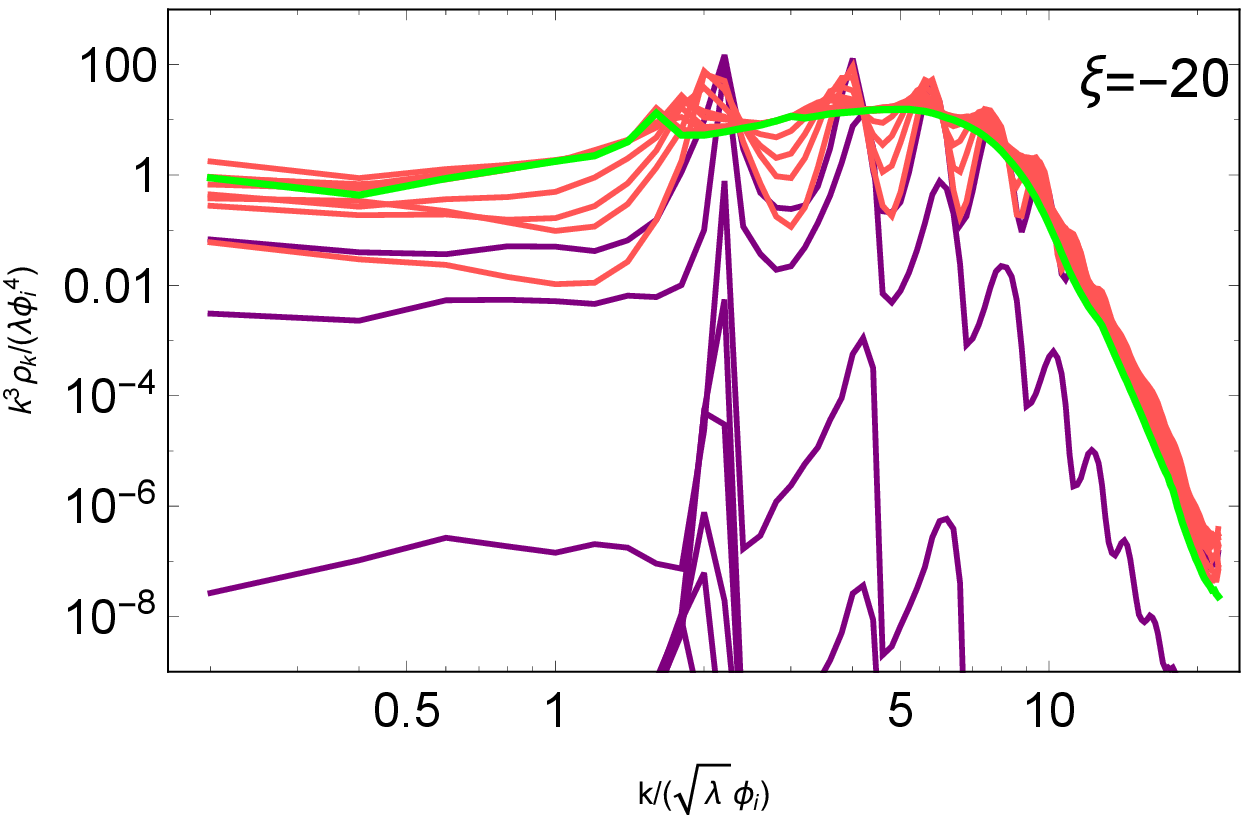}
 \includegraphics[width=0.49\textwidth ]{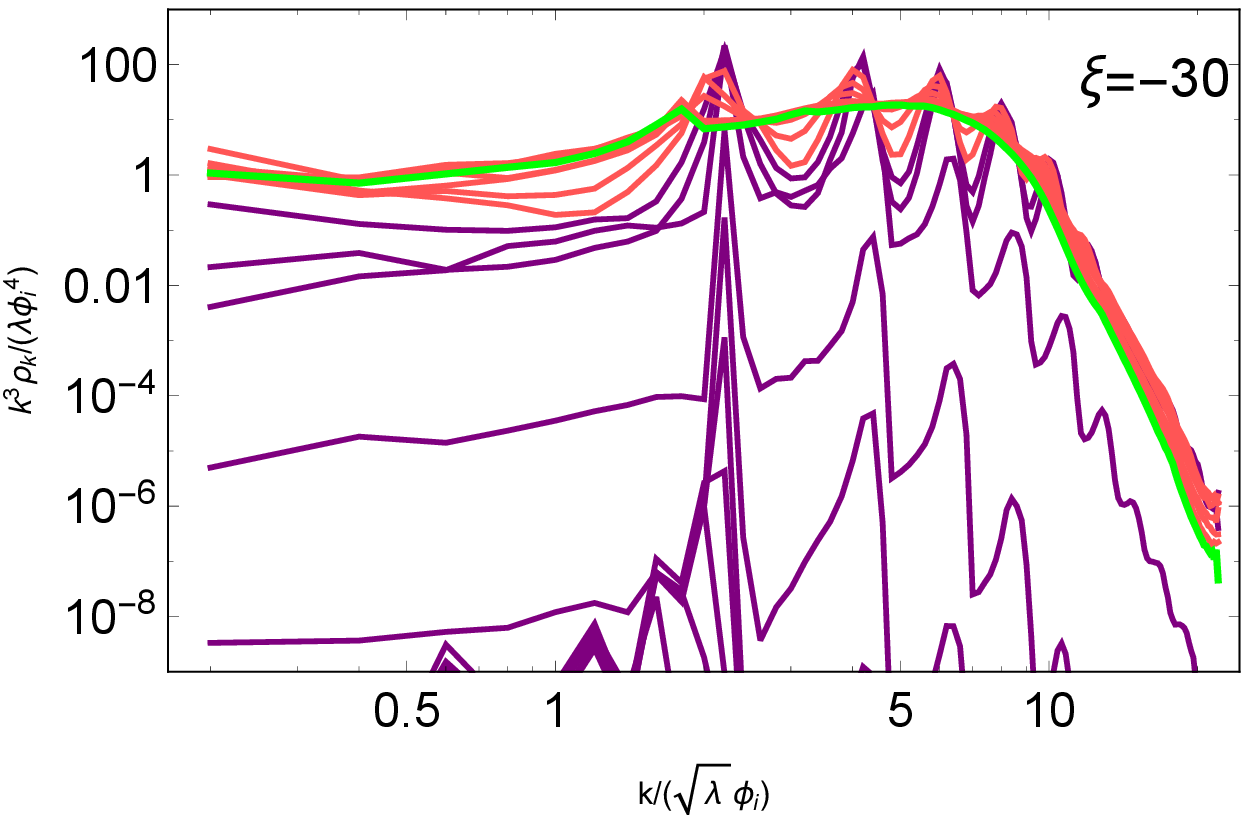}
\caption{\label{fig1} The evolutions of the energy density spectra $k^3\rho_k/(\lambda\phi_i^4)$ of the inflaton quanta  as a function of $k/(\sqrt{\lambda}\phi_i)$ with time $\tau\equiv\eta/(\sqrt{\lambda}\phi_i)$. The spectra from bottom  to up are plotted with the time interval $\Delta\tau=50$, with green line corresponding to the final result. The purple and red lines represent the early and late times, respectively. }
\end{figure}

We show in Fig.~\ref{fig1} the evolution of the energy density spectrum $k^3\rho_k/(\lambda\phi_i^4)$ of the inflaton quanta,  where  $\rho_k=\frac{1}{2}\left(\omega_k^2|\varphi_k|^2+\left|\varphi_k^{\prime}\right|^2\right)$ is the comoving energy density of created momentum modes~\cite{Figueroa2},  as a function of   $k/(\sqrt{\lambda}\phi_i)$ with time $\tau\equiv\eta/(\sqrt{\lambda}\phi_i)$ for different values of  $\xi$ ($\xi=0,-5,-20$ and $-30$). Although the coupling constant is constrained to be $\xi<-1.9\times10^{-3}$ \cite{Planck13, SJS}, we still consider the case of $\xi=0$ as a comparison. One can see that the nonminimal coupling modifies the resonant structure and affects the  amplitude of the density spectrum which increases with the increase of $|\xi|$, but does not change the spectrum shape.    At the initial linear stage of  preheating,    the growth of  inflaton quanta  takes place mainly in the  resonance bands.  This leads to the formation of  the peak structure in the spectra.
  When the backreaction effects of created momentum modes are  significant,  the evolution of the inflaton quanta enters a fully nonlinear stage and then the main peaks of the spectra stop growing and reach their maxima.  Due to the scattering among different modes, the peaks  created during the early stages are gradually washed out.   An interesting feature is that the maximum peak  gradually decreases and moves lightly to $k=0$, but  it does not disappear completely.

 Now we have seen that the perturbations of the scalar inflaton are amplified by the parametric resonance. But,  these perturbations do not cause significant changes on the scalar power spectrum  at  large length scales and hence do not affect the predictions of inflation for the CMB~\cite{Bassett}. On small scales,  the growth of fluctuations may lead to copious production of primordial black holes~\cite{Bassett}. However,  whether the amplitude of the power spectrum of the scalar inflaton perturbations at short length scales  is consistent or not with the bounds from  ultracompact minihalo objects  and primordial black holes~\cite{Emami}   remains  unclear at present. Nonetheless, this is an interesting issue which we would rather leave for future investigation, since it is beyond the scope of the present paper.

\section {Production of gravitational waves}
Now, we study the production of GWs during preheating, which corresponds to the transverse-traceless tensor perturbations $h_{ij}$ of the flat FRW  metric.   In the gravitational theory  with a nonminimal coupling between the scalar field and the curvature scalar,  $h_{ij}$ obeys, to the first order, the following equation of motion
\begin{equation}\label{10}
\ddot h_{ij}+\left(3H+\frac{\dot F}{F}\right)\dot h_{ij}-\frac{1}{a^2}\nabla^2 h_{ij}=\frac{2\kappa^2}{a^2F}\Pi_{ij}^{TT}\;,
\end{equation}
with
\begin{equation}\label{11}
\Pi_{ij}\equiv (1-2\xi)\partial_i\phi\partial_j\phi-2\xi\phi\partial_i\partial_j\phi\;,
\end{equation}
where $F\equiv1-\xi\kappa^2\phi^2$,  $\kappa^2=8\pi G $, and $\Pi_{ij}^{TT}$ is the transverse-traceless part of $\Pi_{ij}$. The term $\Pi^{TT}_{ij}$ associated with the inhomogeneous inflaton field sources the gravitational radiation. In comparison with  the minimally coupled model, we find that  the source of gravitational waves $\Pi_{ij}^{TT}$ contains two extra $\xi$-dependent terms which predominate for $|\xi|>1/2$.  The energy density associated with GWs is given by \cite{CKJ}
\begin{equation}\label{18}
\rho_{gw}=\sum\limits_{i,j}\frac{1}{32\pi G}\langle \dot h^2_{ij}\rangle\;.
\end{equation}
Here $\langle\cdot\cdot\cdot\rangle$  is  the spatial average.

In our numerical calculation,  we introduce  the transverse-traceless projection operator~\cite{Bellido08}:   $\Lambda_{ij,lm}=P_{il}P_{jm}-\frac{1}{2}P_{ij}P_{lm}$, where $
P_{ij}(\mathbf{k})=\delta_{ij}-\frac{k_i k_j}{k^2}$, to obtain the transverse-traceless part of $\Pi_{ij}$   in the momentum space
\begin{equation}\label{12}
\Pi^{TT}_{ij}(\mathbf{k})=\Lambda_{ij,lm}\Pi_{lm}(\mathbf{k}).
\end{equation}
Using this projection operator,  one can define a new tensor $u_{ij}$ which satisfies the following relation in the momentum space
\begin{equation}\label{15}
h_{ij}(\mathbf{k})=\Lambda_{ij,lm}u_{lm}(\mathbf{k})\;.
\end{equation}
Instead of directly investigating the evolution of $ h_{ij}$ in the configuration space, we  solve numerically  the equation of motion of $u_{ij}$,
\begin{equation}\label{16}
\ddot u_{ij}+\left(3\frac{\dot a}{a}+\frac{\dot F}{F}\right)\dot u_{ij}-\frac{1}{a^2}\nabla^2 u_{ij}=\frac{2\kappa^2}{a^2F}[(1-2\xi)\partial_i\phi\partial_j\phi-2\xi\phi\partial_i\partial_j\phi]
\end{equation}
with  $u_{ij}$  and its derivative being initialized as zero.   This  method is different from both the one of solving  Eq.~(\ref{10}) directly~\cite{Easther3} and the one based on the Green's function~\cite{Dufaux}.
Then the energy density (\ref{18}) can be rewritten as
\begin{align}\label{20}
\rho_{gw} =\frac{1}{8GL^3}\int d\ln{k} k^3 \int\frac{d\Omega_k}{4\pi}\Lambda_{ij,lm}\dot u_{ij}(\mathbf{k})\dot u^{\ast}_{lm}(\mathbf{k})\;,
\end{align}
with $L$ the length of one side of the lattice, and  the corresponding spectra of GWs, normalized to the  critical energy density $\rho_c$, can be obtained from \begin{equation}\label{21}
\Omega_{gw}\equiv \frac{1}{\rho_{c}}\frac{ d \rho_{gw}}{d\ln{k}}=\frac{\pi k^3}{3H^2L^3} \int\frac{d\Omega_k}{4\pi}\Lambda_{ij,lm}\dot u_{ij}(\mathbf{k})\dot u^{\ast}_{lm}(\mathbf{k})\;.
\end{equation}

\begin{figure}
\centering
 \includegraphics[width=0.49\textwidth ]{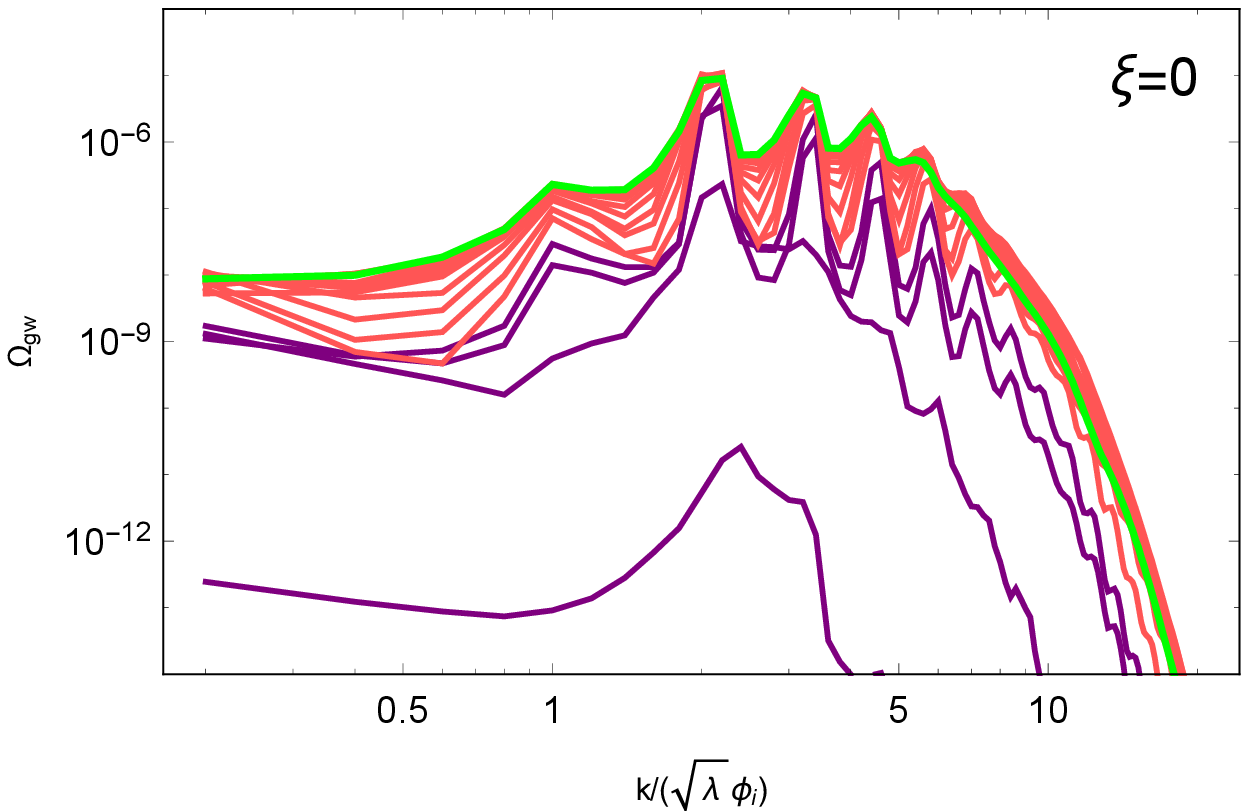}
 \includegraphics[width=0.49\textwidth ]{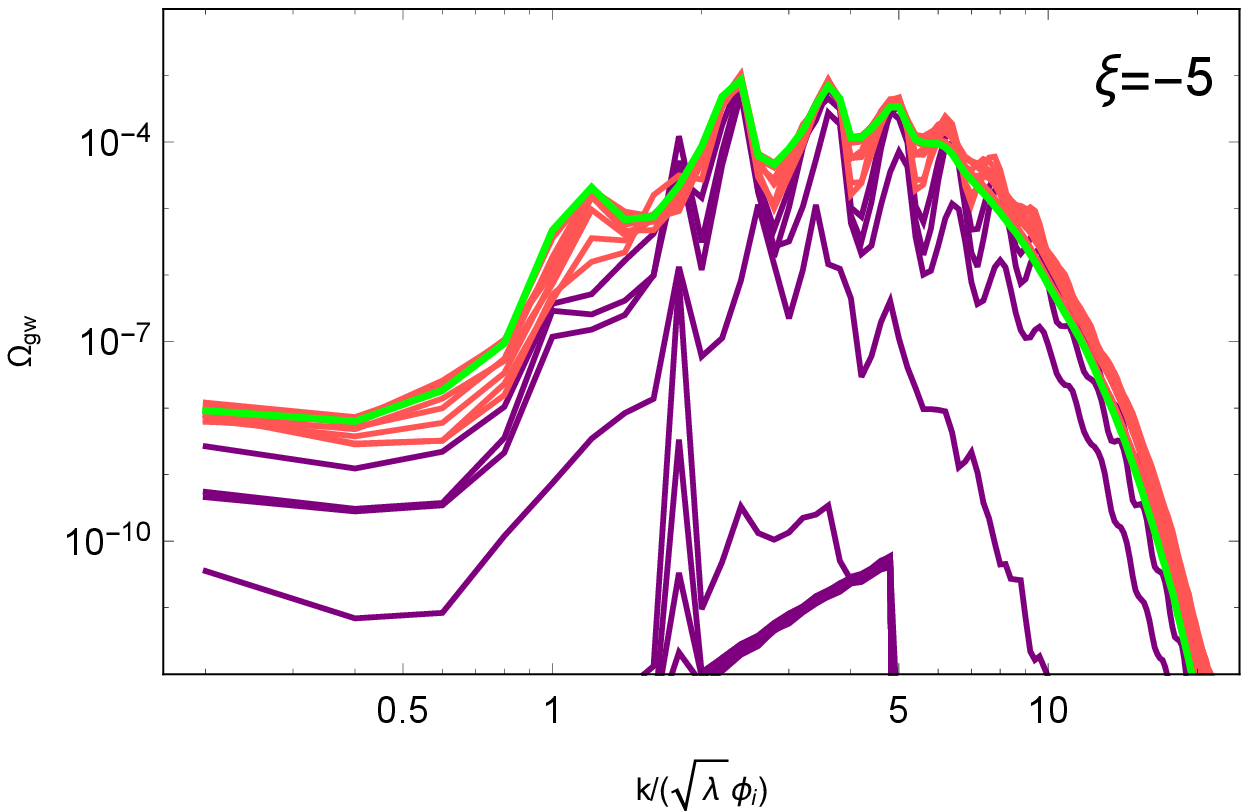}
 \includegraphics[width=0.49\textwidth ]{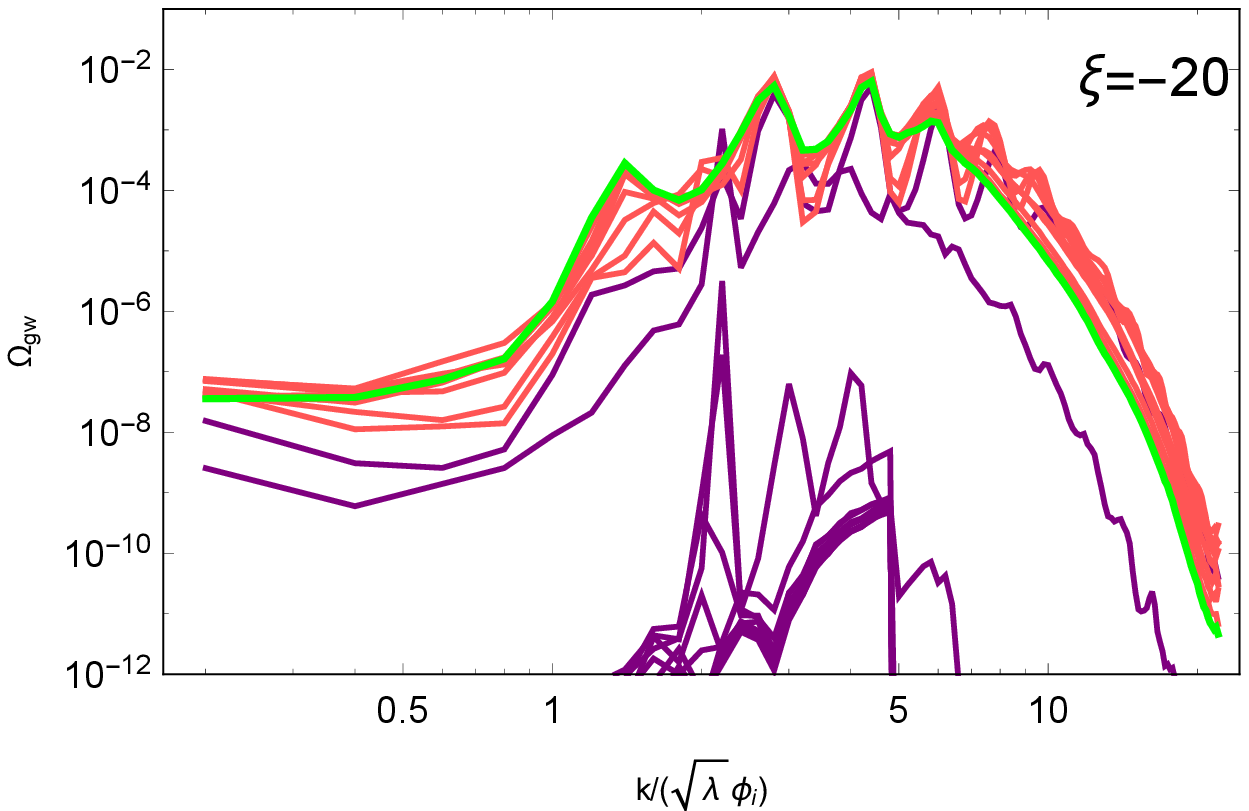}
 \includegraphics[width=0.49\textwidth ]{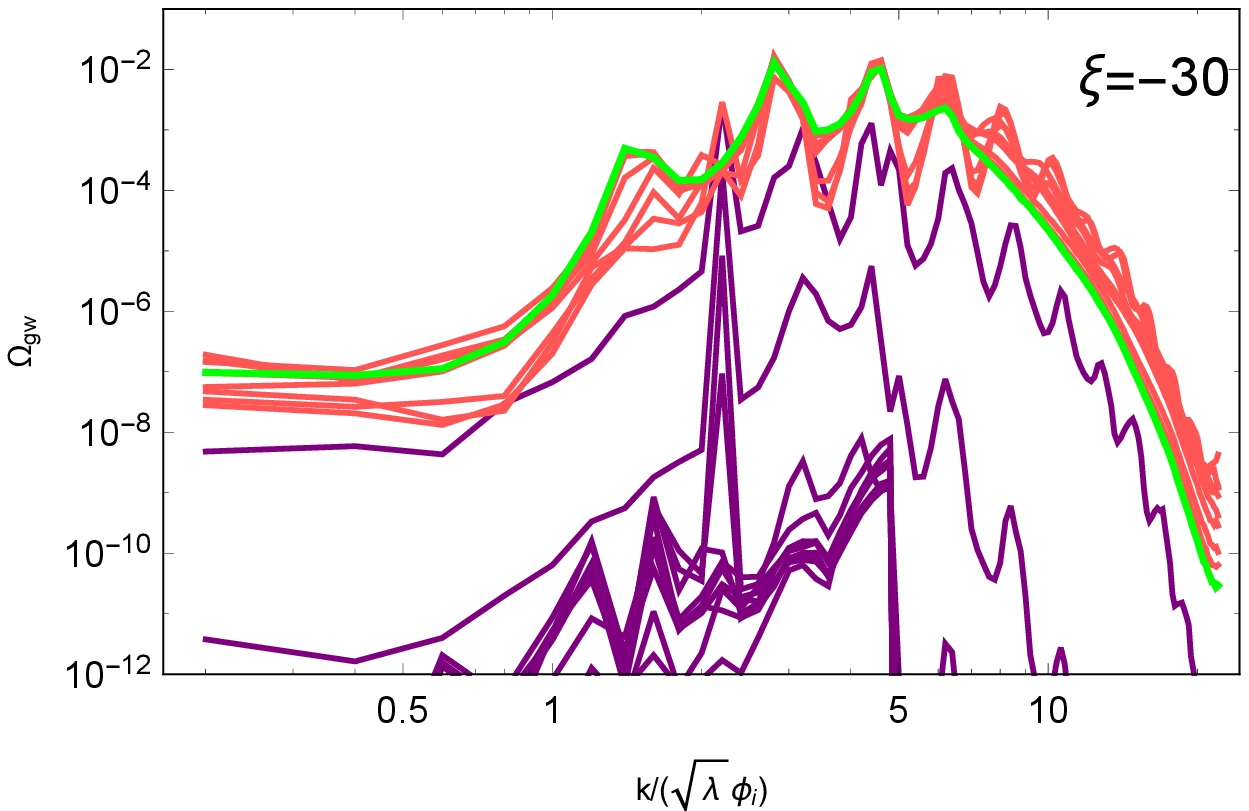}

\caption{\label{fig2} The evolutions of the density  spectra of GWs as a function of  $k/(\sqrt{\lambda}\phi_i)$ with time $\tau$.  The spectra from bottom  to up are plotted with the time interval $\Delta\tau=50$, with green line corresponding to the final result. The purple and red lines represent the early and late times, respectively. }
\end{figure}

The evolutions of the GW density  spectra from the lattice simulation are shown in Fig.~\ref{fig2} with $\xi=0, -5$, $-20$ and $-30$. A paramount characteristic is that the final density spectra of GWs have several distinct peaks. These peaks  are relative to the  resonant bands. Furthermore, the late time evolution of GW spectra leads to a minor peak in the low momentum region, which originates from the non-vanishing  peak of the energy density spectrum of the inflaton quanta.   For the case of a minimal coupling ($\xi=0$), the peak value of  GWs generated in the main resonance band is noticeably  larger than those from  subordinate  bands, which is consistent with the result from the $\phi^4$ chaotic inflation with an interaction between inflaton and  a massless scalar field~\cite{Khlebnikov,Figueroa}.      For the case of $\xi\neq 0$,   one can see that there is a pronounced peak of GWs at the initial stage, which disappears completely in the final spectrum.  Our numerical check indicates that this peak results from the contribution of  the last term on the right hand side of Eq.~(\ref{11}). This term also suppresses the growth of the low-frequency  GWs, and  enhances the high-frequency part, so that the produced GWs  in the subordinate resonance bands become substantial.

With the increase of $|\xi|$,  the peak values of the spectra increase since the energy density amplitude of inflaton quanta is strengthened.   The maximum peak values of the spectrum are $1.57\times10^{-5}, 8.59\times10^{-4}$, $6.32\times10^{-3}$ and $1.30\times10^{-2}$ for $\xi=0,-5,-20$ and $-30$, respectively. The numerical results tell us that the  maximum peak value of the spectrum has a larger value for  a larger   $|\xi|$ and thus the generated GWs will have strong backreactions on the background evolution.  The backreaction effect of GWs, however, is beyond the scope of this paper, so we do not consider  further bigger values of $|\xi|$.

\begin{figure}
\centering
\includegraphics[scale=0.85]{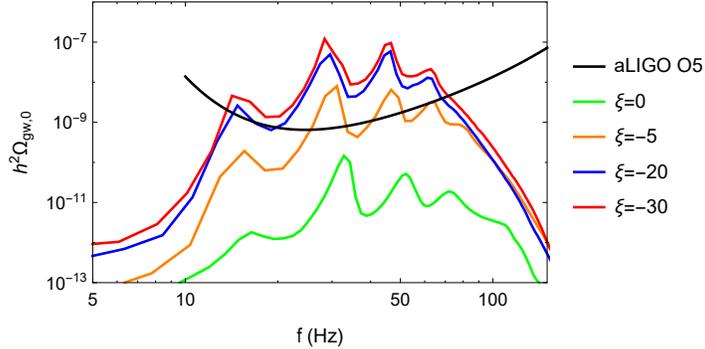}

\caption{\label{fig3} Today's spectra of GWs  with $\lambda=3 \times 10^{-38}$. The black curve is the expected sensitivity curve of the fifth observing run (O5) of the aLIGO-Virgo detector network. }
\end{figure}

Since the model parameter $\lambda$ is constrained by the amplitude of the CMB temperature fluctuations to be $\lambda\simeq4\times10^{-10}\xi^2$ \cite{Bauer, SJS} (when $|\xi \kappa^2 \phi^2 |\gg 1$ during inflation), after projecting the energy spectrum of GWs generated during preheating into today, its peak frequency is beyond $10^7Hz$ and blueshifts with the increase of $|\xi|$. This frequency exceeds  many orders of magnitude the frequency possibly detectable  by current GW experiments.   However, the energy scale of inflation   determines  the peak frequency scale of GWs rather than the peak amplitude of the energy spectrum~\cite{Easther, Easther2}.  Thus, if we relax the constraint on $\lambda$ from the CMB observations, frequencies lying in the present experiments detection range might be possible. For example, setting  $\lambda=3\times10^{-38}$,  the peak frequency is about $30Hz$ when $\xi=-30$.
Since the amplitude of GW spectrum does not reduce,  its spectrum  could lie above the expected sensitivity curve of the fifth observing run (O5) of the aLIGO-Virgo detector network \cite{BRD}.  Fig.~\ref{3} represents our predictions for GW spectra today. One can see that there are two peaks for $\xi=-5$ and four peaks for $\xi=-20$ and $-30$ falling within the range of the O5 detection. Our results are different from that  obtained in~\cite{Antusch, Liu}  where  the pronounced peak is one for the asymmetric potential  and two  for  the cuspy potential.  They   are also different from that of the hybrid preheating~\cite{Bellido} where the amplitude  of GWs is significant for the high-scale model of inflation, but has no apparent peaks. Thus, the detection of GWs in the future will provide us a chance to differentiate different inflationary models, to understand underlying physics of preheating since the GW peaks are related with the parametric resonance, as well as  to obtain an upper limit on the coupling strength  as the GW amplitude is determined by the value of the coupling parameter.

    Now  a few comments are in order for our choice of $\lambda \sim10^{-38}$ to get a GW spectrum, which might  be within the reach of the aLIGO-Virgo detector network, and our neglect of couplings of the inflaton to other fields during the preheating.    First, we  want to  point out  that  a value  of $\lambda$ as tiny as  $ \sim10^{-38}$ may not be  as unrealistic as it appears to be at the first sight.
 For example, it is well known that in the hybrid inflationary scenario~\cite{Linde} $\lambda$ is essentially a free parameter, and   $\lambda \sim10^{-38}$ corresponds to the inflationary energy scale of about $10^7$Gev, which is within the allowed region of a successful model of inflation.
Second, in this paper we only examined the $\delta \phi-$particle production. In addition to this process, the $\phi$ field can also decay to other particles $\chi$ through the interaction, i.e.  $\frac{1}{2}g^2 \phi^2 \chi^2$, but the $\delta\phi-$particle production   appears to be the leading process for $g^2\ll \lambda$ in the minimally coupled case~\cite{Greene}.  So, if the coupling of the inflation field with matter is  very weak, it is reasonable  to assume  these additional couplings can be neglected safely for  the model considered here.
However, in this case, our universe might not  be thermalized to a high enough temperature by the production of   $\delta\phi-$particles. This is because that   the relation between the reheating temperature $T$ and the coupling constant $\lambda$ has been shown to be  $T \propto  \sqrt{\lambda} $  in the case of nonminimal coupling~\cite{Crespo}, and $T\sim 0.1$Gev when $\lambda\sim10^{-38}$. Therefore,  the $g^2 \phi^2 \chi^2$-like couplings   are needed to thermalize our universe.
The  periodic oscillation of $\phi$ field will lead   to the  resonant production of    $\chi$ particles inevitably, which can amplify the reheating temperature. If the universe can be thermalized to a high enough temperature, the standard model thermal plasma can be generated by the self-interaction of $\delta \phi$-particles along with the decay of inflaton field through the $g^2\phi^2 \chi^2$-like interactions.  But   the reheating temperature may not be able to reach a high enough value to allow for big bang baryongenesis when $g^2/\lambda$ is very small. For example,   in the minimally coupled case,  it has been found that  the reheating temperature can  expressed as $ \frac{1}{10} m_{pl}  \lambda^{1/4}(\frac{g^2}{\lambda})^2  \log^{-1} (1/\lambda)$ when the parametric resonance is considered ~\cite{Shtanov}  and   therefore it can not reach  the electroweak energy scale  when $g^2/\lambda <0.01$ and  $\lambda \sim10^{-38}$.   As the reheating temperature  depends on the magnitude of the couplings between the inflaton field and other matter, we also need to consider the resonant productions of both inflaton quanta and $\chi$ particles at the same time.  Moreover,  the reheating also depends on whether  the $\chi$ field   couples  with  gravity or not.    If  the $\chi$ field is  coupled nonminimally with  gravity ($\xi' R \chi^2$), our numerical simulations indicate that the tachyonic preheating  will occur because the effective mass of the $\chi$ field becomes negative when $\phi$ drops below a critical value.
 Since an systematic  analysis of preheating with the couplings of the inflaton field with other matter fields not neglected  is very complicated and  the tachyonic preheating  is different from  the preheating  considered in this paper, we will not go any further on them  but leave to a future paper.

  \section{Conclusion}
We have investigated  the preheating and the in-process
production of GWs after inflation  in which  the inflaton is nonminimally coupled to the curvature in a self-interacting quartic potential. We find that the  amplitude of the density spectrum of inflaton quanta is enhanced by the  nonminimal coupling and it increases with the increase of the coupling parameter.  As a result,  the peak values of the GW spectrum generated during preheating also increase with the increase of the coupling parameter and
might reach the detection limit  of future GW experiments. We also find that the peaks of GW spectrum generated during preheating in which the inflaton is nonminimally coupled to the curvature exhibit  distinctive characteristics as compared to those of minimally coupled inflaton potentials and reflect the story of parametric resonance. Thus,  the detection of these peaks in the future will provide us a new avenue to reveal the physics of  the early universe.

\begin{acknowledgments}
We appreciate very much the insightful comments and helpful suggestions by  an anonymous referee,  and  thank Zong-Kuan Guo and Jing Liu very much for  providing  us  the data for the expected sensitivity curve of the fifth observing run  of the aLIGO-Virgo detector network.  This work was supported by the National Natural Science Foundation of China under Grants No. 11775077,  No. 11435006, No.11690034, and  No. 11375092.
\end{acknowledgments}

\end{document}